\documentclass{aa}
\usepackage{graphicx}

\usepackage{psfig}
\def\gta{ \lower .75ex \hbox{$\sim$} \llap{\raise .27ex \hbox{$>$}} }
\def\lta{ \lower .75ex\hbox{$\sim$} \llap{\raise .27ex \hbox{$<$}} }
\begin{document}



\author{P.O. Petrucci\inst{1} \and
L. Maraschi\inst{2} \and F. Haardt\inst{3} \and K. Nandra\inst{4}}

\offprints{P.O. Petrucci}

\institute{Laboratoire d'Astrophysique de Grenoble, BP 43, 38041 Grenoble
  Cedex 9, France \email{petrucci@obs.ujf-grenoble.fr} \and Osservatorio
  Astronomico di Brera, Via Brera 28, 20121 Milano, Italy
  \email{maraschi@brera.mi.astro.it} \and Dipartimento di Scienze,
  Universit\'a dell'Insubria, via Lucini 11, 22100 Como, Italy
  \email{Francesco.Haardt@mib.infn.it} \and Astrophysics group, Imperial
  College London, Blackett Laboratory, Prince Consort Road, London SW7
  2AZ, UK \email{k.nandra@imperial.ac.uk}}

\date{}

\title{Physical interpretation of the UV/X-ray variability behavior of
NGC 7469}

\abstract{We present a re-analysis of the simultaneous $\sim$ 30 day
  IUE/XTE observation of NGC 7469 done in 1996. Our main progress in this
  paper, in comparison to previous spectral analysis (Nandra et al.
  \cite{nan98,nan00}), is to adopt and fit directly to the data a
  detailed model of the Comptonized spectrum. This firstly allows to fit
  simultaneously the data from the UV to the hard X-ray band in a
  self-consistent way and secondly it gives direct constraints on the
  physical parameters of the disc-corona system, like
  the temperature and optical depth of the corona.\\
  Our results are { completely} consistent with a slab geometry where
  all the observed UV emission is supposed to cross the corona but more
  photon-starved geometries also give acceptable fits. { Whatever the
    geometry is, the UV seed photon emission appears to be dominated by
    the reprocessing of the X-rays.}\\
  We also found very interesting correlations between the different model
  parameters, the most important one being the anticorrelation between
  the corona temperature $kT_e$ and the UV flux $F_{UV}$.  { Such an
    anticorrelation is clearly inconsistent with a fixed disc-corona
    configuration and suggests a variation of the geometry of the
    system.}
  We also find a correlation between the corona optical depth and the
  X-ray flux which may reflect processes linked to the corona formation.\\
  During the observations, NGC 7469 appears to accrete near its Eddington
  limit. This source could then be an example of magnetically dominated
  disc-corona system as recently proposed by Merloni (\cite{mer03}).
  Finally, these data strongly support the presence of a pair free
  corona.  }

\authorrunning{Petrucci et al.}
\maketitle

\section{Introduction}
The broad band opt-UV-X-$\gamma$-ray spectra of Seyfert galaxies is
mainly dominated by two components: an optical/UV bump, which was
interpreted early on early as the signature of cold material accreting
onto a putative super massive black hole (Shields \cite{shi78}; Malkan \&
Sargent \cite{mal82}), and a broader component, covering the soft-X to
the hard-X/soft$\gamma$-ray energy band, with a rough power law shape
generally cut-off above 100 keV.

These spectra are commonly interpreted in the framework of the
reprocessing/upscattering models (e.g. Haardt \& Maraschi, \cite{haa91},
\cite{haa93b}). These models assume the presence of two phases, a cold
(generally an accretion disc) and a hot (called the corona) one,
radiatively linked one with each other: part of the cold emission, giving
birth to the UV bump, is produced by the reprocessing of part of the hot
emission. Inversely the hot emission, at the origin of the broad X-ray
component, is believed to be produced by Compton upscattering of the soft
photons, emitted by the cold phase, on the coronal energetic electrons.

The presence of a high energy cut-off near 100 keV, first detected by
SIGMA and OSSE in NGC 4151 (Jourdain et al. \cite{jou92}; Maisack et al.
\cite{mai93}), appears to be common in Seyfert galaxies (Zdziarski et al.
\cite{zdz93}; Gondek et al. \cite{gon96}; Matt \cite{mat01}; Perola et
al. \cite{per02}). The presence of this component, and the lack of any
annihilation lines (while poorly constrained) are generally believed to
be the signature of the thermal nature of the coronal plasma (see however
Malzac \& Jourdain \cite{mal00}), the cut-off energy being then of the
order of the coronal temperature. No cut-off but strong annihilation
lines were indeed expected by the different non-thermal models develop in
the 80-90s.  However when taking better into account the feedback of the
accelerated particles on the acceleration process (Done et al.
\cite{don90}; Henri \& Pelletier \cite{hen91}; Petrucci, Henri \&
Pelletier al. \cite{pet01}) non-thermal models also predict spectral
shape in
agreement with the observations.\\

Stronger and less ambiguous constraints on the nature of the coronal
plasma are expected from variability studies. For example, in the case of
a thermal hot corona in radiative equilibrium with the cold phase, the
X-ray spectral shape is expected to harden (the photon index $\Gamma$
decreases) when the corona temperature (and thus the high energy cut-off)
increases (Haardt et al. \cite{haa97}). Inversely, in the case of a
non-thermal plasma, a softening of the spectrum is expected when the
cut-off energy increases to ensure the balance between the accelerated
particles and the shocked gas pressures (Petrucci et al. \cite{pet01}).

Such variations of the X-ray spectral shape may be produced by intrinsic
changes of the corona properties (for example changes of the heating
process efficiency) and/or by variation of the external environment like
changes of the soft photons flux (and consequently of the corona cooling)
produced by the cold phase. Simultaneous information in the UV and X-ray
band are then needed to constrain the nature of the different processes
acting in both phases.\\

Several observations have already suggested a link between the UV and/or
the extreme UV and the X-ray emission in Seyfert galaxies (Clavel et al.
\cite{cla92}; Edelson et al. \cite{ede96}; Marshall et al. \cite{mar97};
Chiang et al. \cite{chi00}; Uttley et al. \cite{utt00}; Shemmer et al.
\cite{she01}), showing strong correlations (with small or zero time lag)
between the different fluxes, in good agreement with the
reprocessing/upscattering interpretation. In some cases however, the
X-rays and the optical/UV bands appear to be completely disconnected. For
instance, the $\sim$1 month simultaneous IUE/RXTE monitoring campaign on
NGC 7469 performed in 1996, showed only poor correlations between the
2-10 keV and UV fluxes (Nandra et al. \cite{nan98}, hereafter N98).
Strong X-ray variations on small time scale were also detected but none
in UV. Similar behavior has been observed in NGC 3516 (Edelson et al.
\cite{ede00}; Maoz et al. \cite{mao00,mao02}). These last results ruled
out apparently, and at least for these sources, the basic interpretation
presented above.\\

The spectral analysis of the 1996 RXTE data of NGC 7469 by Nandra et al.
(\cite{nan00} hereafter N00) however permit to resolve these
contradictions for this source.  These authors indeed found that the
X-ray spectral index in NGC 7469 was strongly correlated with the UV flux
at zero time lag during the campaign. They interpret this result as a
strong support for the X-rays being produced by Compton upscattering of
the UV photons. In this framework, the observed variability of the UV
flux is expected to directly modify the cooling rate of the hot corona,
thus producing the observed X-ray spectral changes. The delays found
between the RXTE hard X-rays with respect to the soft ones by Papadakis,
Nandra, \& Kazanas (\cite{pap01}) also offers additional support to the
hypothesis that the X-ray are most likely due to Comptonization of soft
photons by hot, thermal electrons. The apparent disconnect between the
X-ray and UV fluxes previously reported by N98, was then explained by the
narrow band used in the X-ray (2-10 keV), which, in the presence of
spectral variability, may result in incoherent
behavior with respect to the UV one.\\

The spectral treatment done by N00 assumed a simple power law shape for
the X-ray continuum. It is known however that the real Comptonization
spectral shape may be significantly different from a power law (Haardt
\cite{haa93}; Svensson \cite{sve96}; Petrucci et al.  \cite{pet00}).
Moreover, the interpretation of the spectral variability in term of
physical quantities, like the temperature and optical depth of the
corona, is not straightforward with a phenomenological power law
approximation and requires the use of approximate formulae. The aim of
this paper is then to perform a re-analysis of the simultaneous IUE/XTE
data by explicitly fitting the UV and X-ray continuum using a realistic
Comptonization model. As just said, the first interest of this
re-analysis is that we will have direct information on physical
quantities characterizing the hot and cold phases. The broad band (from
UV to hard X-rays) energy range of the Comptonization spectra also permit
to fit simultaneously the IUE and RXTE data and to predict the spectral
behavior at high energy (i.e. above 20 keV, the upper limit of our RXTE
data) in a completely
consistent way.\\

The paper is organized as follows. We first present the data and the
model used in Sect. 2. We then show, in Sect. 3, the results we obtain
fitting simultaneously the IUE and XTE data and we also present the most
interesting correlations we found between the different model parameters.
We discuss these different results in Sect. 4 and then conclude on this
work in the last section.

\section{Observation and data analysis}
\subsection{The data}
Many of the details of the RXTE observations are reported in N98 and N00
and the reader can refer to these papers for more details.  The
``L7+activation'' model was employed for the background. The standard
selection criteria in the ``rex'' RXTE processing script has also been
applied.

The background subtraction was however unreliable at high energies for
sources as weak as NGC 7469 and the spectral analysis was therefore
restricted to the 2-20 keV energy band. Different response matrices were
calculated for different times of the observation using PCARSP v2.37 but
no significant differences were found between the derived spectra. We
therefore used a single matrix for the entire observation.

The IUE data used here are described in Wanders et al. (\cite{wan97}).
NGC 7469 has also been observed by the HST FOS instrument in the midst of
the IUE monitoring (Kriss et al. \cite{kri00}). The better quality of the
HST data has allowed to disentangle the UV continuum more accurately from
the broad wings of the emission lines and thus to identify four clean
continuum windows free of any emission and/or absorption features, and
centered on 1315, 1485, 1740 and 1825 \AA. We fit the data using
XSPECv11.1. Since the XSPEC fitting tools need spectral bin larger than 1
eV, we have rescaled the UV fluxes in order to have 4 windows of 1eV band
width. In the following, and to be comparable with the results of Wanders
et al. (\cite{wan97}) and Kriss et al. (\cite{kri00}), the UV fluxes are
given in erg.s$^{-1}$.cm$^{-2}$.\AA$^{-1}$ and are measured in a bandwidth
of 21 \AA.

Like in N00, the observation have been divided into 30 segments of
approximately 1 day duration with sufficient photon statistics to permit
precise spectral analysis.

\subsection{The model}
\label{model}
We fit the data using, for the continuum, the thermal Comptonization
model in slab geometry developed by F.  Haardt (Haardt \cite{haa93};
Haardt \cite{haa94}; Haardt et al.  \cite{haa97}). Hereafter we will
refer to this Anisotropic Comptonization Code as AC2. { The code
  treats Comptonization in steady-state, i.e., energy balance is
  implicitly assumed, so that an unspecified heating term balances
  radiative losses.}  Given the temperature and optical depth of the
corona, the angle--dependent spectra of a disc--corona system in plane
parallel geometry is derived using an iterative scattering method, where
the intrinsic anisotropy of the scattering is taken into account only in
the first scattering order.  The Compton recoil is also treated
accurately.  It includes also a reflection component described following
White, Lightman \& Zdziarski (\cite{whi88}) and Lightman \& White
(\cite{lig88}) and assuming neutral matter. The spectral shape of the
reflected photons is averaged over angles. It is multiplied by a first
normalization factor which depends on the inclination angle (see
Ghisellini, Haardt \& Matt \cite{ghi94} for details). In addition, the
usual $R$ normalization is left free to vary in the fit procedure, so
that, for the given inclination angle, $R=1$ corresponds to a solid
angle, subtended by the reflector, of 2$\pi$.  { The resulting
  Comptonization spectral shape generally is quite different from the
  simple cut-off power law approximation generally used to mimic
  Comptonization spectra. Moreover, due to anisotropy effects, the first
  scattering order is reduced in comparison to the other orders, and the
  observed spectral shape is better approximated by a broken power law,
  the energy of the break depending mainly on $kT_e$ and $kT_{bb}$. It is
  expected to be in the 2-10 keV range for the parameter values we used
  (cf. Petrucci et al.  \cite{pet00} for more details).}

The fit parameters in AC2 are the temperature of the corona $kT_e$, its
optical depth $\tau$, the temperature of the disc $kT_{bb}$ (assumed as a
single temperature black body) and the reflection normalization $R$.
There are then two extra parameters compared to the simple power law +
reflection model used by N00.  Other natural outputs of AC2 are the total
(integrated over the solid angles) X-ray fluxes emitted by the corona
(upward toward the observer and backward toward the disc) as well as the
total UV flux emitted by the disc and the total flux Compton reflected at
the disc surface.  Different grids of this model are now available for
public use as XSPEC table models\footnote{From
  http://pitto.mib.infn.it/~haardt/ATABLES}.

{ It is important to point out that no links between the Comptonized
  spectrum and the soft UV disc emission are imposed a priori.  The model
  simply adjusts its parameters to fit the data. It is only a posteriori
  that the resulting best fit values can be interpreted in a physically
  motivated scenario. More specifically, the code gives the ratio between
  an arbitrarily normalized input cooling flux, $F_{cool}$, and the total
  emitted Compton spectrum, $F_{X}$. That ratio can then be related to a
  specific geometry of the system, e.g. a homogeneous plane parallel
  disc-corona system in radiative equilibrium as in Haardt \& Maraschi
  (\cite{haa91}), in which case it must be $F_{cool}\simeq F_{X}$.  This
  balance is imposed by the feed-back of the corona on the disc and
  viceversa in the absence of a dissipation mechanism intrinsic to the
  disc and is independent on the heating rate, provided it heats the
  electrons in the corona. A doubling of the heating rate doubles $F_{X}$
  and $F_{cool}$ {\it but does not change their ratio}. If $F_{X}\gta
  F_{cool}$, a photon starved configuration is required for instance a
  spherical corona. In the most general case, $F_{cool}$ is a fraction of
  the observed UV flux $F_{UV}$. In fact, the condition $F_{UV}=F_{cool}$
  is strictly valid only if all UV photons cross (and hence cool) the
  corona. In general $F_{UV} \geq F_{cool}$.  Thus, once the best fit
  parameters are obtained, one should seek for a self-consistent (in
  terms of physics and geometry) configuration of the disc+corona
  system.}

AC2 also does not treat the ionization and thermal balance of the cold
phase at all. This has certainly some impact on the broad band spectral
shape expected from such a model as shown by Malzac et al.
(\cite{mal03}), leading, for instance, to higher corona temperature and
harder spectra in comparison to the ``blackbody + neutral reflection''
approximation used here. We believe however that these effects have no
strong influence on the spectral variability behaviors studied in this
paper.

We have also the possibility of using a semi-spherical geometry for the
corona. We have checked however that the results presented below do not
significantly change in comparison to the slab configuration (the main
change being the larger value of the optical depth for a given slope of
the X-ray power law). We thus restrict our analysis to the slab geometry.

We used the {\sc uvred} model for the reddening, fixing the extinction to
the best fit value of Kriss et al. (\cite{kri00}) i.e. $E(B-V)=0.12$. We
fixed the column density to the galactic one i.e. $N_{H}=4.8\times
10^{20}$cm$^{-2}$. For the neutral iron line, we simply used a gaussian
({\sc zgauss} model of {\sc xspec}), fixing the gaussian energy and width
to 6.4 keV and 0.01 keV respectively. The gaussian normalization was let
free to vary. Finally we assume a relative normalization of 1 between the
IUE and XTE data (we have checked that normalizations of 0.9 or 1.1 do
not significantly change the results).

\begin{figure}
\includegraphics[width=\columnwidth]{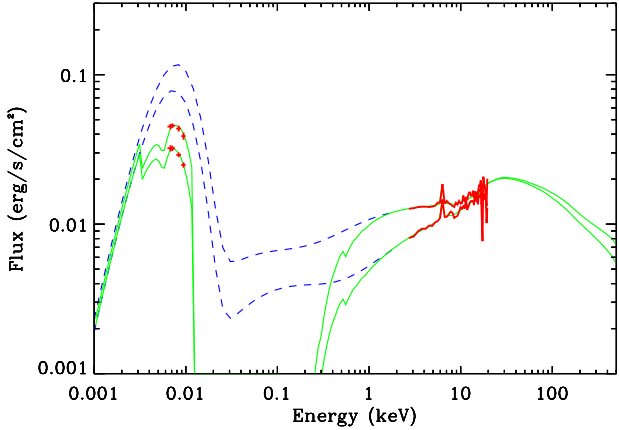}
\caption{Data and best fit model corresponding to a soft (segment 2) and
  hard (segment 9) state of NGC 7469. The data are in thick (red) lines,
  the unfolded model (including the reddening and the neutral hydrogen
  absorption) in thin solid (green) lines, and the unfolded model
  (without reddening or neutral hydrogen absorption) in dashed (blue)
  lines.}
\label{figplotfit}
\end{figure}

\section{Simultaneous IUE and XTE fits}
\subsection{Fit results}
\begin{figure}[h!t]
\includegraphics[width=\columnwidth]{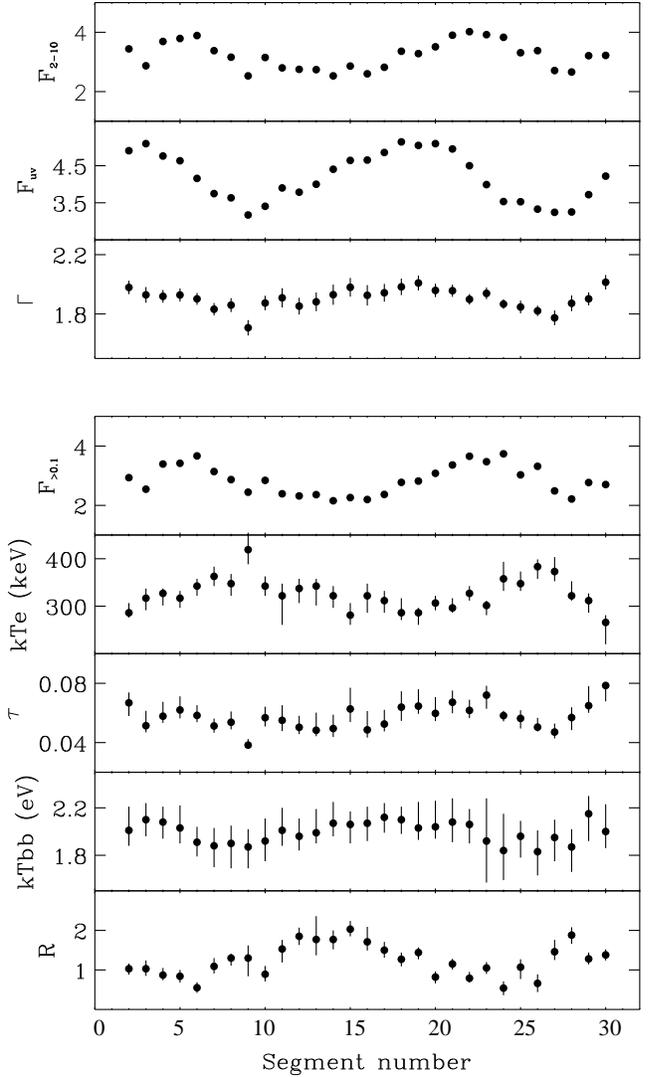}
\caption{The light curves of the different parameters of the model
  (i.e. $kT_e$, $\tau$, $kT_{bb}$ and $R$) are plotted in the bottom part
  of the figure as well as the total (above 0.1 keV) X-ray flux predicted
  by the AC2 code (in 10$^{-10}$ erg.s$^{-1}$.cm$^{-2}$). For comparison,
  we have also plotted, at the top, the observed UV flux at 1315\AA (in
  10$^{-14}$ erg.s$^{-1}$.cm$^{-2}$.\AA$^{-1}$), the 2-10 keV X-ray flux
  (in 10$^{-11}$ erg.s$^{-1}$.cm$^{-2}$) and the photon index light curve
  obtained by N00.}
\label{lcpara}
\vspace*{-0.5cm}
\end{figure}
We have reported in Table \ref{tab1} the best fit values of the different
model parameters, i.e. $kT_e$, $\tau$, $kT_{bb}$ and $R$, for 29 daily
segments fitting simultaneously the IUE and XTE data. We do not include
the first segment obtained at the beginning of the campaign,
JD-24400245.104, since no IUE data were available. We have also reported
in the table
the UV (at 1315 \AA) and X-ray (in the 2-10 keV range) fluxes, and the
$\chi^2$ values.
We have plotted in Fig. \ref{figplotfit} the data and the corresponding
best fit models for two different states of NGC 7469 (a soft one
corresponding to segment 2 and a hard one corresponding to segment 9).\\

\begin{table*}
\begin{center}
\begin{tabular}{ccccccccc}
\hline
Day & $kT_e$ & $\tau$ & $kT_{bb}$ & $R$ & $F_{2-10}$ &
$F_{1315}$ & $\chi^2$ \\
(TDJ+24400000) & (keV) & & (eV) & & ($\times 10^{-11}$) & ($\times 10^{-14}$) &
(46 dof)\\
\hline
246.177 & 286$_{-10}^{+20}$ & 0.067$_{-0.009}^{+0.007}$ & 2.0$_{-0.1}^{+0.2}$ & 1.0$_{-0.1}^{+0.1}$ &   3.44 & 4.90 & 50.0 \\
247.247 & 317$_{-26}^{+20}$ & 0.051$_{-0.004}^{+0.010}$ & 2.1$_{-0.1}^{+0.1}$ & 1.0$_{-0.2}^{+0.2}$ &   2.87 & 5.09 & 49.3 \\
248.316 & 327$_{-26}^{+10}$ & 0.058$_{-0.004}^{+0.010}$ & 2.1$_{-0.1}^{+0.1}$ & 0.9$_{-0.1}^{+0.2}$ &   3.69 & 4.76 & 62.4 \\
249.386 & 317$_{-20}^{+15}$ & 0.062$_{-0.006}^{+0.009}$ & 2.0$_{-0.1}^{+0.2}$ & 0.8$_{-0.2}^{+0.2}$ &   3.79 & 4.63 & 67.6 \\
250.456 & 342$_{-20}^{+15}$ & 0.058$_{-0.004}^{+0.007}$ & 1.9$_{-0.1}^{+0.1}$ & 0.6$_{-0.1}^{+0.1}$ &   3.89 & 4.15 & 34.5 \\
251.579 & 363$_{-20}^{+20}$ & 0.051$_{-0.004}^{+0.005}$ & 1.9$_{-0.2}^{+0.2}$ & 1.1$_{-0.2}^{+0.2}$ &   3.38 & 3.74 & 65.3 \\
252.631 & 347$_{-26}^{+20}$ & 0.054$_{-0.005}^{+0.007}$ & 1.9$_{-0.2}^{+0.1}$ & 1.3$_{-0.2}^{+0.1}$ &   3.16 & 3.63 & 41.3 \\
253.676 & 419$_{-31}^{+36}$ & 0.038$_{-0.003}^{+0.004}$ & 1.9$_{-0.2}^{+0.1}$ & 1.3$_{-0.5}^{+0.3}$ &   2.53 & 3.17 & 50.4 \\
254.714 & 342$_{-20}^{+20}$ & 0.057$_{-0.006}^{+0.007}$ & 1.9$_{-0.2}^{+0.2}$ & 0.9$_{-0.2}^{+0.2}$ &   3.15 & 3.40 & 45.0 \\
255.777 & 322$_{-61}^{+26}$ & 0.055$_{-0.007}^{+0.010}$ & 2.0$_{-0.1}^{+0.2}$ & 1.5$_{-0.3}^{+0.2}$ &   2.80 & 3.90 & 62.0 \\
256.811 & 337$_{-31}^{+20}$ & 0.050$_{-0.005}^{+0.008}$ & 2.0$_{-0.1}^{+0.1}$ & 1.9$_{-0.2}^{+0.2}$ &   2.75 & 3.78 & 45.2 \\
257.840 & 342$_{-41}^{+15}$ & 0.048$_{-0.004}^{+0.012}$ & 2.0$_{-0.1}^{+0.2}$ & 1.8$_{-0.4}^{+0.6}$ &   2.74 & 4.00 & 43.9 \\
258.865 & 322$_{-26}^{+20}$ & 0.050$_{-0.006}^{+0.009}$ & 2.1$_{-0.1}^{+0.2}$ & 1.8$_{-0.3}^{+0.2}$ &   2.53 & 4.40 & 49.9 \\
259.914 & 281$_{-20}^{+26}$ & 0.063$_{-0.009}^{+0.014}$ & 2.1$_{-0.2}^{+0.1}$ & 2.0$_{-0.2}^{+0.2}$ &   2.86 & 4.64 & 43.4 \\
260.914 & 322$_{-36}^{+26}$ & 0.049$_{-0.005}^{+0.013}$ & 2.1$_{-0.1}^{+0.1}$ & 1.7$_{-0.2}^{+0.4}$ &   2.60 & 4.65 & 56.6 \\
261.918 & 312$_{-26}^{+20}$ & 0.053$_{-0.005}^{+0.010}$ & 2.1$_{-0.1}^{+0.1}$ & 1.5$_{-0.2}^{+0.2}$ &   2.82 & 4.85 & 42.8 \\
262.882 & 286$_{-15}^{+31}$ & 0.064$_{-0.009}^{+0.011}$ & 2.1$_{-0.1}^{+0.1}$ & 1.3$_{-0.2}^{+0.2}$ &   3.36 & 5.14 & 56.3 \\
263.883 & 286$_{-26}^{+10}$ & 0.065$_{-0.005}^{+0.011}$ & 2.0$_{-0.1}^{+0.2}$ & 1.4$_{-0.2}^{+0.1}$ &   3.28 & 5.04 & 55.8 \\
264.917 & 307$_{-15}^{+15}$ & 0.060$_{-0.005}^{+0.011}$ & 2.0$_{-0.1}^{+0.2}$ & 0.8$_{-0.2}^{+0.1}$ &   3.51 & 5.09 & 40.6 \\
265.953 & 296$_{-10}^{+20}$ & 0.067$_{-0.007}^{+0.008}$ & 2.1$_{-0.2}^{+0.2}$ & 1.2$_{-0.1}^{+0.1}$ &   3.90 & 4.95 & 64.9 \\
267.019 & 327$_{-15}^{+15}$ & 0.062$_{-0.005}^{+0.007}$ & 2.1$_{-0.2}^{+0.1}$ & 0.8$_{-0.1}^{+0.2}$ &   4.02 & 4.50 & 61.8 \\
268.084 & 301$_{-20}^{+10}$ & 0.072$_{-0.009}^{+0.006}$ & 1.9$_{-0.4}^{+0.4}$ & 1.1$_{-0.1}^{+0.2}$ &   3.92 & 3.99 & 53.8 \\
269.155 & 358$_{-26}^{+36}$ & 0.058$_{-0.004}^{+0.003}$ & 1.8$_{-0.3}^{+0.3}$ & 0.5$_{-0.2}^{+0.2}$ &   3.83 & 3.53 & 58.4 \\
270.222 & 347$_{-15}^{+26}$ & 0.056$_{-0.007}^{+0.006}$ & 2.0$_{-0.2}^{+0.1}$ & 1.1$_{-0.3}^{+0.2}$ &   3.31 & 3.52 & 43.8 \\
271.282 & 383$_{-26}^{+15}$ & 0.050$_{-0.003}^{+0.006}$ & 1.8$_{-0.2}^{+0.2}$ & 0.7$_{-0.2}^{+0.2}$ &   3.38 & 3.32 & 58.0 \\
272.361 & 373$_{-20}^{+31}$ & 0.047$_{-0.004}^{+0.006}$ & 2.0$_{-0.2}^{+0.1}$ & 1.5$_{-0.2}^{+0.3}$ &   2.71 & 3.24 & 59.8 \\
273.417 & 322$_{-10}^{+31}$ & 0.057$_{-0.008}^{+0.007}$ & 1.9$_{-0.2}^{+0.1}$ & 1.9$_{-0.2}^{+0.2}$ &   2.66 & 3.25 & 62.5 \\
274.488 & 312$_{-26}^{+15}$ & 0.065$_{-0.005}^{+0.013}$ & 2.2$_{-0.2}^{+0.1}$ & 1.3$_{-0.1}^{+0.2}$ &   3.21 & 3.71 & 41.4 \\
275.547 & 266$_{-46}^{+15}$ & 0.079$_{-0.011}^{+0.009}$ & 2.0$_{-0.1}^{+0.2}$ & 1.4$_{-0.1}^{+0.1}$ &   3.22 & 4.21 & 61.7 \\
\hline
\end{tabular}
\end{center}
\caption{Best fit values of our model parameters $kT_e$, $\tau$,
  $kT_{bb}$ and $R$ obtained when fitting the XTE and IUE data
  simultaneously. We have also reported the X-ray and UV (at 1315 \AA) 
  fluxes (in erg.s$^{-1}$.cm$^{-2}$ and erg.s$^{-1}$.cm$^{-2}$.\AA$^{-1}$
  respectively)} 
\label{tab1}
\end{table*}
We obtained acceptable fits for each segment and the total $\chi^2$ =
1528 for 1334 dof.  The use of the IUE and XTE data provide relatively
good constraints on the different parameters of the model. Indeed, the UV
data allow to constrain well $kT_{bb}$. Moreover, in Comptonization
models, and for a given geometry, the ratio between the UV bump maximum
and the X-ray ``plateau'' { (i.e. the X-ray flux near 0.1 keV)} is
mainly controlled by the corona optical depth. Combined with the X-ray
slope in the 2-10 keV range it then permits relatively good determination
of $\tau$ and $kT_e$ and thus of the X-ray continuum up to the hard
X-ray/soft $\gamma$-ray band.  Consequently, the reflection component is
also well constrained even if the XTE data permit precise spectral study
only below 20 keV.\\

\subsection{Light curves of fit parameters}
We have plotted the light curves of the different model parameters on the
bottom part of Fig. \ref{lcpara}, as well as the total (above 0.1 keV)
X-ray flux predicted by the AC2 code. We have also plotted the 2-10 keV
X-ray and UV (at 1315 \AA) light curves and the photon index one,
obtained by N00, on the top of the figure.

We have tested the significance of the variability of each parameter.
Fitting the different light curves with a constant gives a $\chi^2$, for
29 dof, of 57.6, 54.9, 8.3 and 128.5 for $kT_e$, $\tau$, $kT_{bb}$ and
$R$, respectively. Except $kT_{bb}$, the other parameters are thus
clearly variable (significant at $>$99\% confidence).

A more sensitive way to detect parameters variations is the use of the
F-test.  For that purpose, we fit again the data but fixing each of the
parameters ($kT_e$, $\tau$, $kT_{bb}$ and $R$) to its weighted average
value (i.e. 322 keV, 0.054, 2 eV and 1.1 respectively). By comparing the
total $\chi^2$ of these fits to the $\chi^2$ when all parameters were
free, we derived F-values of 4.3, 3.9, 1.2 and 4.9 for 29 additional
parameters. Again, it implies highly significant changes in $kT_e$,
$\tau$ and $R$ ($>$99\%) but no significant changes for $kT_{bb}$. The
black body temperature $kT_{bb}$ is thus consistent with a constant and
we choose to fix it, in the following, to the mean value of 2 eV.

We finally remark that the total X-ray flux (above 0.1 keV) predicted by
our model also varies similarly to the observed 2-10 keV flux.



\subsection{Correlations}
\begin{table*}[!ht]
\begin{center}
\begin{tabular}{ccccccccc}
\hline
& $kT_e$ &  $\tau$  & R  & F (2-10)  & F$_{1315}$  & F$_{line}$  \\
\hline
$kT_e$ & ... &   {\bf -0.65}  & -0.26  & -0.15  & {\bf -0.73}  & -0.03 \\
$\tau$ & {\bf -0.79}  & ... & -0.30  & {\bf0.67}  & 0.37  & 0.23 \\
R & -0.22  & -0.21  & ... & {\bf -0.79}& -0.05  & -0.31  \\
F (2-10) & -0.20  & {\bf 0.59} & {\bf -0.79} & ... & 0.35  & 0.40 \\
F$_{1315}$ & {\bf -0.67}  & 0.35  & -0.07  & 0.30 & ... & -0.05 \\
F$_{line}$ & -0.16  & 0.33  & -0.34  & 0.43  & 0.04  & ... \\
\hline
\end{tabular}
\end{center}
\caption{The Spearman (rank) correlation coefficients are reported in the
upper right portion of the Table and the Pearson linear ones in the lower
left part. The correlations have 29 points. The correlation coefficients
with formal chance probabilities less than 1\% (i.e. with correlation
coefficient $>$ 0.46) are written in bold
faces.}
\label{paracorrel}
\end{table*}
We have reported in Table \ref{paracorrel} the results of the linear
(Pearson) and rank (Spearman) correlation tests between the different
model parameters, assuming no time lags.

\begin{figure}[b]
\includegraphics[width=\columnwidth]{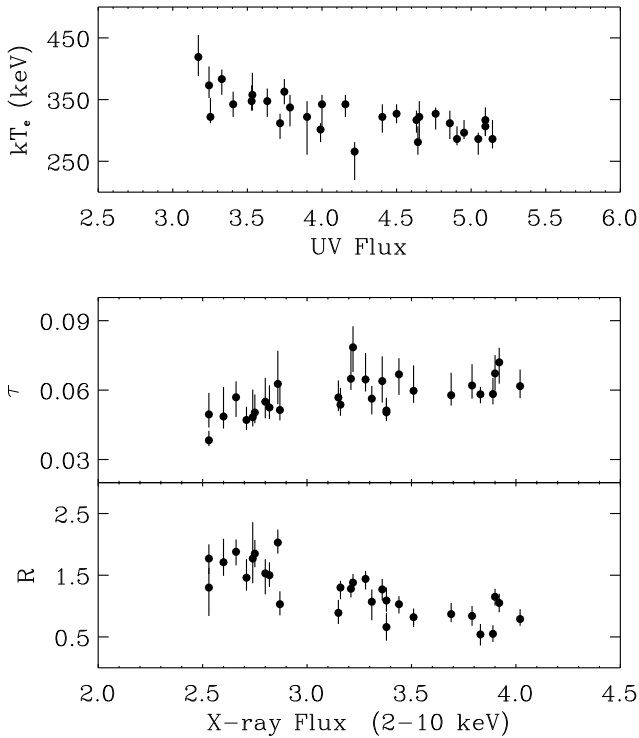}
\caption{Top: coronal temperature
  vs. UV flux at 1315\AA (in erg.s$^{-1}$.cm$^{-2}$.\AA$^{-1}$). Bottom:
  coronal optical depth $\tau$ and reflection normalization $R$ vs. 2-10
  keV flux (in erg.s$^{-1}$.cm$^{-2}$). }
\label{correlplot}
\end{figure}

The X-ray flux is apparently correlated, with significances $>$99\%, with
the coronal optical depth $\tau$ and anti-correlated with $R$. This
anticorrelation was not observed by N00. We also found a correlation
between the X-ray and line fluxes but not as strong as N00.  Concerning
the UV flux, it appears strongly anti-correlated with the coronal
temperature $kT_e$. Finally, we found a strong anti-correlation between
$kT_e$ and $\tau$. We have plotted the most significant correlations in
Figs.  \ref{correlplot} and \ref{tauvstheta} i.e $\tau$ and $R$ versus
$F_{2-10}$, $kT_e$ versus $F_{1315}$, and $kT_e$ versus $\tau$.

As already remarked by N00, the cross correlation of light curves that
have a ``red noise'' character (i.e. where the variability power scales
as $f^{-\alpha}$), can produce artificially high correlation coefficients
values (Welsh \cite{wel99}; Maoz et al. \cite{mao00}). We have thus
tested the significance of our results by simulating a number of X-ray
and UV light curves with ``red noise'' power spectra. We adopt a value of
$\alpha$ of 1.3 and 1.9 for the power law index of the X-ray and UV
power-density spectra respectively (cf. N98 and N00).

In 200 trials of 29 points, we never obtained Spearman or Pearson
correlation coefficients as high as that observed in the real data.
%
We thus conclude that the correlations shown in bold in Table
\ref{paracorrel} are unlikely to arise by chance.  We discuss these
different correlations in the following.

\subsubsection{$kT_{e}$ versus $F_{UV}$}
The coronal temperature appears strongly anticorrelated with the UV flux
{ ($F_{1315}$ mimics very well the total UV flux variations predicted
  by our model and in the following we will now only use the term
  $F_{UV}$ instead of $F_{1315}$)}.  We found a rank Spearman coefficient
of 0.73 while the highest coefficient obtained with our simulations,
using 200 trials, was $\sim$ 0.5 (0.6 for 400 trials). This
anti-correlation clearly corresponds to the $\Gamma$ vs.  $F_{UV}$
correlation found by N00.

These authors interpret this correlation as a strong support for
Comptonization models { where any "primary" increase of the UV seed
  photons (not accompanied by an increase of the coronal heating rate) is
  generally expected to produce an increase of the cooling of the corona
  and thus a steepening of the spectrum and a decreasing of the corona
  temperature.  We will see in the following that, in the context of
  coupled disc-corona models where the UV arises from reprocessing, a
  geometrical and/or energetic change of the disc-corona configuration is
  required. }


\subsubsection{$\tau$ versus $F_{2-10}$}
\label{tauFx}
We find a strong correlation between the coronal optical depth and the
X-ray emission. { As already noted in Sect. \ref{model}, in the code
  we used there is no link imposed a priori between $\tau$ and the total
  X-ray flux $F_{X}$ and consequently between $\tau$ and $F_{2-10}$}.
Such correlation is not simply explained in the Comptonization model
framework and may reflect a process of different nature.

For instance, we expect that part of the X-ray flux produced by the
corona illuminates the surrounding cold thick material, seed of soft
photons. It is also possible that the hot corona is produced by e.g.
evaporation of part of this cold matter or by magnetic buoyancy inside
the cold matter (Meyer \& Meyer-Hofmeister \cite{mey94}; Meyer et al.
\cite{mey00}; Liu et al.  \cite{liu02}; Merloni \& Fabian \cite{mer02};
Merloni \cite{mer03}). Then variations of the X-ray flux may likely
modify the properties of the upper layers of the accretion disc, and thus
influence the corona formation. For instance, thermal instabilities are
known to exist in illuminated plasma (R\'o\.za\'nska \& Czerny
\cite{roz96}; Nayakshin, Kazanas \& Kallman \cite{nay00}), and the degree
of ionization and temperature of the disc surface may rapidly increase in
response to an increase of the illuminating flux. It may then favor the
evaporation of part of the disc matter, increasing consequently the
corona optical depth (but see Sect. \ref{edd}).

\subsubsection{$kT_e$ vs $\tau$}
\label{tevstau}

\label{tauvsthetasect}
\begin{figure}[h!t]
\includegraphics[width=\columnwidth]{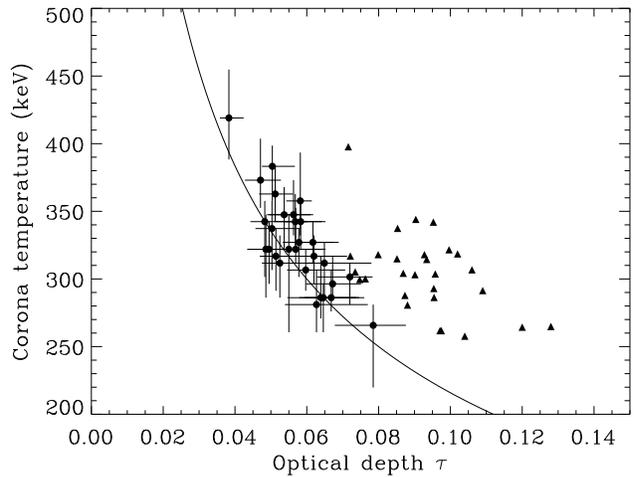}
\caption{Coronal temperature vs. coronal optical depth. Filled circles
  (with errors) have been obtained assuming that all the observed UV flux
  has crossed the corona. Filled triangles correspond to the case where
  only half of the observed UV flux is supposed to have crossed the
  corona. The errors have been omitted in this case for clarity. The
  relation expected in the case of a slab corona above a passive disc has
  been overplotted in solid line (cf. Sect. \ref{tevstau} for details).}
\label{tauvstheta}
\end{figure}
We have reported on Fig. \ref{tauvstheta} the coronal optical depth
$\tau$ versus the coronal temperature $kT_e$ deduced from our fits.  A
strong anti-correlation is clearly visible. Such anti-correlation is
indeed expected in the case of a corona-disc configuration in radiative
equilibrium. It depends however on the real geometry of the system
(Haardt \& Maraschi, \cite{haa93b}; Stern et al. \cite{ste95}; Malzac et
al. \cite{mal01}). The relation expected in the case of a slab corona
above a passive disc has been overplotted in solid line in Fig.
\ref{tauvstheta}. The data seem to be in good agreement with the
theoretical predictions, even if they slightly stand above the solid
line { and present some dispersion}.\\

It is worth noting that in our fitting procedure, we have assumed that
all the observed UV emission, whatever its origin, has crossed and cooled
the corona.  It is however possible that part of the disc emission does
not interact with the corona, like in the case of a patchy or spherical
corona.  We have simulated such a configuration by adding a black body
component to the model used to fit the data. Its temperature was fixed to
2 eV and its normalization was fixed so that the corresponding flux (at
1315\AA) was half the observed mean UV flux. In this way, we impose that
a fixed part of the UV emission does not cross and cool the corona. The
corresponding values of $\tau$ and $kT_e$ have been overplotted with
triangles in Fig. \ref{tauvstheta}. We omit the error bars for clarity.
The triangles are clearly above the solid line, as expected for a more
photon starved configuration. The total $\chi^2$ that we obtained with
this model is 1592 still for 1334 dof, slightly larger than without the
addition of a black body component, but statistically equivalent. A
geometry more photon-starved than the simple slab corona is thus also in
agreement with the observations.


It is worth noting that the addition of a constant black body component
does not substantially modify the results presented above. In particular,
we still found correlation coefficients of the same order of magnitude
than those reported in Table \ref{paracorrel}.


\subsubsection{$F_{K_{\alpha}}$, $R$ versus $F_{2-10}$}
%
%
We found only a weak correlation between the line flux and the X-ray flux
(cf. NOO who found a stronger correlation) but we did find an
anti-correlation, not noted by N00, between $R$ and $F_{2-10}$.

%

It is known that the measurement of the reflection component depends on
the assumed model for the continuum. For example, large differences are
expected between a simple power law shape and a real Comptonization model
(Petrucci et al. \cite{pet01b}). It may then explain the difference
between our results and those of N00.  It does not explain however why
$R$ is anticorrelated with the 2-10 keV flux. Moreover these results are
relatively difficult to explain in the standard view since both the line
and the reflection component are expected to arise from the same region
and thus to vary in the same way.

It is possible that part of the reflection components is emitted close to
the central region while the other part may be produced by a remote
reflector (indeed the presence of a distant reflector may explain the
$R$-$F_{2-10}$ anticorrelation). In this case, and in the presence of
flux and spectral variability, like for NGC 7469, the measurement of $R$
and $F_{line}$ may not be very reliable and meaningful (Malzac \&
Petrucci \cite{mal02}) especially with the limited band pass and spectral
resolution of XTE and the increasing noise at high energy (near 20 keV).
In agreement with N00, we thus believe that the derived values of $R$ and
$F_{line}$ may not be truly representative of the strength and spectral
shape of the real reflection components in NGC 7469. We will thus not
discuss this parameter in the following.

\section{Discussion}
\label{discuss}

{ Important results have emerged from this analysis.  Firstly, we
  obtained very acceptable fits, from the UV to the X-ray range, of the
  29 daily spectra of NGC 7469 using a realistic thermal Comptonization
  model. This model is able to explain the spectral shape of the
  different UV/X-ray daily spectra but also the UV and X-ray flux
  variability.  In agreement with the conclusion of N00 (see also Chiang
  \cite{chi02}), the absence of a clear correlation between the UV and
  X-ray fluxes, as reported by N98, was thus clearly misleading.  It
  simply results from the complex flux and spectral variability of this
  source which may produce, when observed in narrow energy bands,
  relatively unexpected behaviors. Berkley et al. (\cite{ber00}) have
  failed to reproduce the UV and X-ray light curves of NGC 7469 using a
  simple model where an X-ray point source illuminates an infinite
  accretion disc. This model does not consider spectral changes at all
  and thus cannot reflect the intrinsic spectral complexity of
  comptonization models. We believe that this
  is the main reason of the results they obtained.\\
  We also note at this point that our best fit results reported in Table
  \ref{tab1} are quantitatively different from those obtained by Chiang
  (\cite{chi02}) fitting the same data. We also believe that the
  differences are certainly due to the approximate model used by this
  author to mimic thermal Comptonization spectra. For example, for corona
  optical depth as small as the ones he found (cf. his Fig. 5), a thermal
  Comptonization spectrum is formed of well separated bumps of different
  scattering orders and a simple cut-off power law is a really poor
  approximation.
  
  Secondly, we found very interesting correlations between the UV/X-ray
  fluxes and the model parameters, the most important one being the
  anticorrelation between the UV flux and the corona temperature. It is
  the equivalent of the $F_{UV}$-$\Gamma$ correlation found by N00,
  fitting the data with a simple cut-off power law model.  This
  anticorrelation can be simply explained in the framework of thermal
  Comptonization models. Indeed, the increase of the UV soft photons
  flux, if "primary", generally means an increase of the coronal cooling.
  This then produces a decrease of the corona temperature and a softening
  of the spectrum as observed.
  
  However, important results stem from this interpretation.  First, the
  decrease of the corona temperature due to the increase of the coronal
  cooling is completely true in the case of a pair-free corona. Indeed,
  for a pair-dominated corona, the increase of the cooling would
  correspond to an {\it increase} of the corona temperature rather than a
  decrease (Ghisellini \& Haardt, \cite{ghi94}). As discussed in Sect.
  \ref{compact}, there are actually strong indications supporting the
  presence of a pair-free corona in NGC~7469, thus justifying the above
  interpretation.
  
  More importantly, we note that the $F_{UV}$-$kT_e$ anticorrelation
  cannot be reconciled with a {\it fixed} disc-corona configuration in
  radiative balance. As explained in Sect. \ref{model}, such
  configuration corresponds to a {\it fixed} ratio $F_X/F_{UV}$.  But
  there is a priori no link between $F_{UV}$ and the corona parameters
  like its temperature or optical depth. $kT_e$ and $\tau$ have only to
  adjust themselves in order to insure a constant heating/cooling ratio
  (cf. Sect.  \ref{model}). There is thus no reason for $kT_e$ and/or
  $\tau$ to change, whatever the value of $F_{UV}$.
  
  
  As discussed in the following sections, the $F_{UV}$-$kT_e$
  anticorrelation necessarily suggests a change of the geometry of the
  disc-corona configuration.}



\subsection{UV emission dominated by reprocessing}
In the Comptonization code we used, there is no constraint, {\it a
  priori}, on the origin of the UV emission. It may be dominated by the
intrinsic disc emission, by the reprocessed one, or both may be of the
same order. A check can be made {\it a posteriori} by comparing the
$\tau$-$kT_e$ values with the theoretical expectations as we have done in
Sect. \ref{tevstau} and Fig. \ref{tauvstheta}.  If the intrinsic disc
emission were dominant, the corona should be overcooled in comparison to
a corona above a passive disc, and the $\tau$-$kT_e$ values should be in
the lower right part of Fig. \ref{tauvstheta}, below the
solid line, which is not what we obtain.\\

\begin{figure}
\includegraphics[width=0.9\columnwidth]{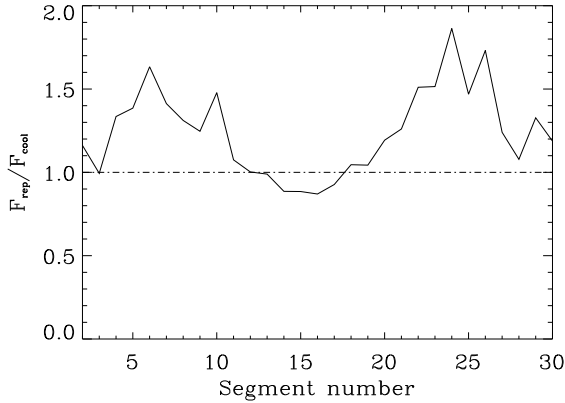}
\caption{Ratio of X-ray flux expected to be reprocessed by the disc,
  $F_{rep}$ and the disc flux which crosses and actually cools the corona
  $F_{cool}$ (solid line).  The theoretical prediction in the case of a
  slab corona above a passive disc in radiative equilibrium is plotted in
  dashed line.}
\label{repvsuv}
\end{figure}

Another way to see this is to compare the X-ray flux expected to be
reprocessed by the disc, $F_{rep}$, and the UV flux which crosses and
effectively cools the corona $F_{cool}$. The former is simply the
difference between the flux emitted by the corona backward toward the
disc and the flux Compton reflected by the disc. For the latter, { and
  in the case of a slab corona completely covering the disc, $F_{cool}$
  is equal to the total disc emission (this does not correspond to the
  {\it observed} UV emission, since part of the disc emission is
  comptonized in the corona).}  Both $F_{rep}$ and $F_{cool}$ are natural
outputs of the Comptonization model used to fit the data.  We have
plotted in Fig.  \ref{repvsuv}, the corresponding ratio
$F_{rep}/F_{cool}$.
{ This ratio appears to be generally larger than 1.  It is also
  variable between 1 and 1.8 and this point is discussed in the next
  section.}
If some part of the UV emission crossing the corona was intrinsic to the
disc we would expect this ratio to be smaller than 1 which is not the
case. 

These  results thus suggest that, in the case of NGC 7469, {\it
  the intrinsic disc emission is negligible in comparison to the
  reprocessed one}. This conclusion is in complete agreement with that of
N00. It also agrees with the observed delays, increasing with wavelength,
observed between different UV bands and interpreted as the different
light travel time between the illuminating X-ray source and the disc
reprocessing regions (Wanders et al. \cite{wan97}; Kriss et al.
\cite{kri00}).

Still concerning Fig.  \ref{repvsuv}, we note that, starting from the
epoch (segm n. 15) where the ratio $F_{rep}/F_{cool}$ is nearly 1, as
expected for the slab configuration, this ratio reaches its maxima in
near correspondence with the maxima of the X-ray light curve (near minima
for the UV fluxes).  One can thus interpret the "oscillation" as a change
in geometry whereby the fraction of reprocessed photons which re-enter
and cool the corona varies.


\subsection{Corona geometry variations as the origin of the observed
  variability}
{ Simulations of Malzac \& Jourdain (\cite{mal01}) show that the time
  scale for a disc-corona system to reach radiative equilibrium, after
  rapid radiative perturbation in one of the two phases, is of the order
  of a few corona light crossing times i.e. well smaller than a day. We
  can thus (reasonably) assume that each of the 29 daily average spectra
  and fluxes of the observation of NGC 7469 correspond to a disc-corona
  system where hot and cold phases are in radiative balance.
  
  We have also shown in the previous section that the disc emission is
  apparently dominated by the X-ray reprocessing.  In this case, a fixed
  disc-corona configuration in radiative equilibrium should correspond to
  a constant $F_X$/$F_{UV}$ ratio (cf. Sect. \ref{model}) contrary to
  what we observe in Fig.  \ref{repvsuv}. We are thus bring to the
  conclusion that {\it the disc-corona geometry necessarily varies during
    the total observation}, producing the observed spectral and flux
  variability.\\

  
  It is not clear how the geometry really varies especially since we have
  no real constraint on the geometry itself.  However, the fact that the
  X-ray emission is apparently the primary source of radiation (the UV
  being only produced by reprocessing), but also of variability (the
  total X-ray flux, above 0.1 keV, is indeed predicted to vary along the
  observation, cf.  Fig. \ref{lcpara}), naturally suggests that the
  change of the disc-corona geometry is actually a change of the corona
  geometry: the larger/smaller the corona, the stronger/weaker the X-ray
  emission.

  We then propose the following scenario.
  Suppose that the origin of the day time scale variability is indeed a
  change of the global corona geometry. Suppose for example that the
  corona, which covers initially a large part of the disc, then becomes
  more patchy. Then the more patchy the corona, the hotter the mean
  corona temperature becomes and the harder the X-ray spectrum is. At the
  same time, the corona covering factor decreasing, the reprocessed flux,
  and consequently the global UV flux, decreases.  This scenario thus
  simply explains the observed $kT_e$ -$F_{UV}$ anticorrelation.
  
  Moreover we may naturally expect the optical depth of the corona to
  depend on the disc X-ray illumination (which mainly occurs below the
  corona blobs) as already suggested in Sect. \ref{tauFx}. The variation
  of the X-ray flux impinging on the disc could likely modify the
  structure of the upper layers of the accretion disc and thus have an
  impact on the corona formation process. A strong X-ray emission would
  then help the evaporation of the accretion disc material thus
  increasing the corona optical depth. This could also explain the
  correlation between the corona optical depth and the X-ray flux.
  
  We of course do not assert that this scenario is unique. We easily
  imagine that the change of the disc-corona geometry is certainly more
  complex, and may also imply changes of the disc itself. For instance,
  its inner region may become geometrically thick and optically thin, due
  to thermal instabilities (however see next section), or its inner
  radius may vary with time.  Consequently, it would modify the UV flux
  entering the corona as well as the reflection component. The time scale
  for such processes to occur is however expected to be quite long in
  comparison to what we observe. Anyway, whatever the scenario proposed,
  the present analysis strongly requires a change of the disc-corona
  configuration. On the other hand, this appears to be a very plausible
  origin for the flux and spectral variability observed in NGC 7469.}

\subsection{A magnetically dominated corona in NGC 7469?}
\label{edd}
The mass of the black hole supposed to be present in the central region
of NGC 7469 has been estimated in between $10^6$ and $10^7$ solar masses.
Precise measurements, using different methods, favor a mass of about
7$\times 10^6 M_{\sun}$ (Collier et al. \cite{col98}; Wandel et al.
\cite{wan99}).  This corresponds to an Eddington luminosity
$L_{edd}\simeq 9\times 10^{44}$erg.s$^{-1}$.

Our model gives us an estimate of the total luminosity emitted by the
central engine in NGC 7469, from the UV to the hard X-rays (if we admit
that the UV emission is in large part dominated by the reprocessing of
X-rays, it is equal to the downscattered emission emitted by the corona).
Assuming a distance of 70 Mpc (with $H_0$=75 km.s$^{-1}$.Mpc$^{-1}$), the
luminosity varies between 2 and 3$\times 10^{44}$erg.s$^{-1}$ i.e. of the
order of $L_{edd}$, meaning that the source accretes near its Eddington
rate.\\

It is known that for such high accretion rate, the inner part of the
accretion disc, which is dominated by the radiation pressure, is
viscously and thermally unstable (Lightman \& Eardley \cite{lig74};
Shakura \& Sunyaev \cite{sha76}) and may be subject to violent clumping
instabilities (Begelman \cite{beg01}; Turner, Stone \& Sano
\cite{tur02}). If the disc remains geometrically thin, most of the
disc mass form very dense clumps embedded in a tenuous hot corona. In
this case however, we expect very soft spectra unless the covering
fraction of the cold clouds is small which results however in low
reflection component (Malzac \& Celotti \cite{mal02}). This does not
agree with the value of $\Gamma$ and $R$ observed in NGC 7469.\\

A possible solution could be the presence of disc-corona configurations
where a major fraction of the accretion power is released in the corona,
as those proposed by Svensson \& Zdziarski (\cite{sve94}), Merloni \&
Fabian (\cite{mer02}), and more recently by Merloni (\cite{mer03}). The
presence of such strong coronae indeed prevents the development of disc
instabilities.  Merloni (\cite{mer03}) found for example new thermally
and viscously stable optically thick accretion disc solutions for
magnetized turbulent flows.  Assuming that the magnetic field
amplification, via MRI, is balanced by buoyant escape, they obtained high
viscosity solutions where a large fraction of the accretion power is
liberated in the corona.  These solutions appear only above a critical
accretion rate of the order of few tenths of the Eddington one, which
corresponds to the case of NGC 7469.  The disc geometry being conserved
in these solutions, we also expect the reflection normalization to be of
the order of unity as observed.\\ 

If this interpretation is correct, it is likely that parts of the disc
involved in the corona formation process are intrinsically weakly
luminous, most of the energy being released mechanicly/magneticaly rather
than radiatively. The observed UV flux would then be dominated by the
reprocessed emission in agreement with the observations.  Moreover, it
could also explain why the black body temperature $kT_{bb}$ deduced from
our fits is significantly smaller than the inner temperature (at $\sim$
10 Schwarzschild radii) of $\sim$ 50 eV expected for a standard accretion
disc (i.e. without corona) assuming the black hole mass and accretion
rate estimates of Collier et al.  (\cite{col98}) from optical data
analysis.



\subsection{A pair free corona}
\label{compact}
As already said, the code we use can give us the total X-ray and UV
fluxes (and consequently the luminosity) emitted by the corona and the
disc respectively. Now, from the variability in both bands, we can also
estimate an upper limit of the size of the emitting regions, $R_{X}$ and
$R_{UV}$, and consequently a lower limit on the compactnesses
$\displaystyle l_{X,UV}=\frac{\sigma_t L_{X,UV}}{m_ec^3R_{X,UV}}$.

\begin{figure}[h!]
\includegraphics[width=\columnwidth]{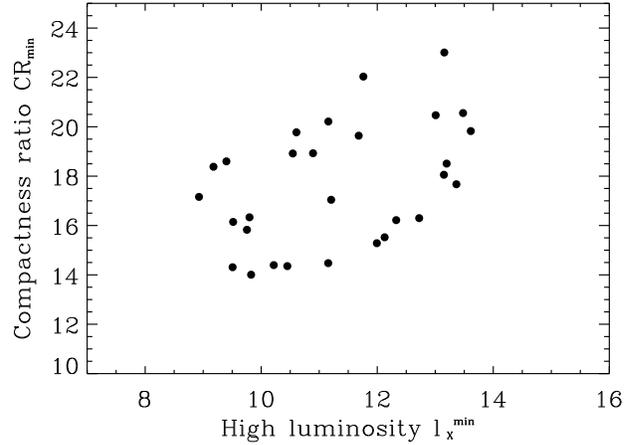}
\caption{The compactness ratio $\displaystyle
  CR_{min}=\frac{l_X^{min}}{l_{UV}}$ versus the X-ray compactness
  $l_X^{min}$ obtained assuming a 1 light day for the X-ray region size
  (cf. Sect. \ref{compact} for details)}
\label{lhlsvslh}
\end{figure}

The X-rays appears to be variable on a timescale of less than 1 day (cf.
N98, N00) thus $\displaystyle l_X > l_X^{min}=\frac{\sigma_t L_X}{R_1
  m_ec^3}$ where $R_1$ is a 1 light day length. On the other hand the UV
vary on a longer time scale of the order of 5 days (Wanders et al.
\cite{wan97}; N98; N00; Collier \& Peterson \cite{col01}). Then the
compactness ratio $CR=\displaystyle \frac{l_X}{l_{UV}} > CR_{min} =
\frac{l_X^{min}}{l_{UV}}\simeq 5\displaystyle\frac{L_X}{L_{UV}}$. We have
reported in Fig.  \ref{lhlsvslh}, the $\displaystyle\left(CR_{min},
  l_X^{min}\right)$ values corresponding to our 29 daily observations. We
obtained a slight correlation between these 2 variables. We also note
that they only vary by about 50-60\% during the campaign.

These results clearly disagree with the pair-dominated expectations
computed by Ghisellini \& Haardt (\cite{ghi94}). Their Fig. 2 shows that,
if the corona of NGC 7469 is pair dominated, its compactness ratio has to
decrease and its high compactness to increase (both by a factor $\sim$
10) from the softest state ($\Gamma$=2.01, $kT_e$=266 keV) to the hardest
state ($\Gamma$=1.7, $kT_e$=420 keV) reach by the source during the 29
days of observation. It would thus require very large changes (of factor
5 or more) of the characteristic size of both the UV and X-ray emitting
regions for our results to be compatible with the pair dominated case.
This is in contradiction with the observations. For instance, the
observed UV flux varying by only 50-60\% imposes the size of the emitting
region (assumed to be optically thick) to vary at most by the same
amount.





\section{Conclusion}
We have re-analyzed in this paper the simultaneous 30 days IUE/XTE
observation of NGC 7469 done in 1996, adopting and fitting directly to
the data a detailed model of the Comptonized spectrum. This contrasts with
previous spectral analysis where a simple power law was used to model the
continuum. We were thus able to fit simultaneously the data from the UV
to the hard X-ray band in a self-consistent way and to obtain direct
constraints on the physical parameters of the disc-corona system, like the
temperature and optical depth of the corona.\\

The model we use is able to explain well the spectral shape of the
different UV/X-ray spectra but also the UV and X-ray flux variability.
The absence of a clear correlation between the UV and X-ray fluxes, as
reported by N98, was thus clearly misleading. It underlines the
importance of the use of consistent models when fitting so intimately
related energy bands. In this context, the observed delay of $\sim$ 0.14
day between the UV and X-ray light curves of NGC 4051, as reported by
Mason et al. \cite{mas02}, may not correspond to a response delay
between the UV and X-ray emitting region.\\
Concerning NGC 7469, the data appear to be roughly consistent with a slab
geometry where all the power is liberated inside the corona and where the
UV flux is dominated by the reprocessing. We note however that more
photon-starved geometries also agree with the data.\\

We also find very interesting correlations between the different model
parameters as the one between the corona temperature and the UV flux.  If
it can be easily explained in the framework of thermal comptonization
model (an increase of the UV flux means an increase of the corona cooling
and thus, in general, a decrease of the corona temperature), it
necessarely implies slight changes of the disc-corona configuration
during the 29 days of observation. We thus propose a model where the main
origin of the variability is a change of the global corona geometry. The
more patchy it becomes, the hotter the temperature. In the same time, the
reprocessed flux, i.e. the UV emission, decreases.\\
We also found an interesting correlation between the corona optical depth
and the X-ray flux which is, a priori, not expected by Comptonization
models and may result from a different, but connected, process linked to
the corona formation.\\



We show that NGC 7469 is apparently accreting at nearly the Eddington
rate. The presence of a reflection component of the order of unity and of
a relatively hard spectrum suggest the presence of a magnetically
dominated disc-corona system in this source where most of the accreting
energy is released in the corona. Such solutions have recently been shown
to be thermally and viscously stable by Merloni (\cite{mer03}). We
finally found strong evidences in favor of a pair free
corona.\\

We note finally that the lack of precise data above 20 keV prevents
however any detailed spectral study of the soft $\gamma$ emission of the
source. For instance, the study of the variability of the high energy
cut-off, would have given additionnal constraints on the model
parameters, especially on the corona temperature. Such very broad band
study would require simultaneous observations with high energy satellites
like XMM and INTEGRAL.


%


\end{document}